# Segregative phase separation scenario of the formation of the bacterial nucleoid

Marc Joyeux[*]

**Abstract :** The mechanism responsible for the compaction of the genomic DNA of bacteria inside a structure called the nucleoid is a longstanding but still lively debated question. Most puzzling is the fact that the nucleoid occupies only a small fraction of the cell, although it is not separated from the rest of the cytoplasm by any membrane and would occupy a volume about thousand times larger outside from the cell. Here, by performing numerical simulations with coarse-grained models, we elaborate on the conjecture that the formation of the nucleoid may result from a segregative phase separation mechanism driven by the demixing of the DNA coil and non-binding globular macromolecules present in the cytoplasm, presumably functional ribosomes. Simulations performed with crowders having spherical, dumbbell or octahedral geometry highlight the sensitive dependence of the level of DNA compaction on the dissymmetry of DNA/DNA, DNA/crowder, and crowder/crowder repulsive interactions, thereby supporting the segregative phase separation scenario. Simulations also consistently predict much stronger DNA compaction close to the jamming threshold. Moreover, simulations performed with crowders of different sizes suggest that the final density distribution of each species results from the competition between thermodynamic forces and steric hindrance, so that bigger crowders are expelled selectively from the nucleoid only at moderate total crowder concentrations. This work leads to several predictions, which may eventually be tested experimentally.

[*] Laboratoire Interdisciplinaire de Physique, CNRS and Université Grenoble Alpes, Grenoble, France. *E-mail* : marc.joyeux@univ-grenoble-alpes.fr

1.   Introduction

The mechanism leading to the formation of the bacterial nucleoid is a longstanding question, which is still lively debated.[1-5] The nucleoid is the region of the cell that contains the genomic DNA, as well as a certain number of proteins and other macromolecules.[6] Its volume varies according to several factors, including the richness of the nutrient,[7-11] the cell cycle step,[12,13] and the action of antibiotics,[9-11,14-18] but it is generally of the order of 25% of the volume of the cell.[19] This is precisely the point that has kept scientists puzzled for decades, because it is estimated (for example, from the Worm Like Chain model[20]) that the volume of the unconstrained genomic DNA of bacteria in physiological solution is approximately thousand times larger than the volume of the cell. Moreover, recent micro-piston experiments have shown that the free energy required to compress the chromosome of *E. coli* cells to its *in vivo* size is of the order of $10^5 \, k_B T$.[21] Since in bacteria, as in all prokaryotes, the DNA is not separated from the main part of the cytosol by a bounding membrane, one is then left with the question of why the DNA does not expand throughout the cell but remains instead localized in the nucleoid.

The mechanisms, which are commonly evoked to explain the formation of the nucleoid, include (i) the formation of plectonemes, (ii) the bridging of DNA duplexes by nucleoid proteins, and (iii) the action of short-range attractive forces,[1] but their actual importance remains still unclear.[4] More precisely, there are corroborating indications that the formation of plectonemes resulting from negative supercoiling leads only to mild compaction of the DNA.[4,22] Physiological concentrations of nucleoid proteins able to bridge two DNA duplexes and keep them at a short distance from one another are, moreover, too low to provoke strong compaction.[4,23] Similarly, the ability of long cationic polymers to shrink progressively the DNA to compact coils that resemble the bacterial nucleoid has been demonstrated *in vitro*,[24] but bacterial cells do not contain significant amounts of such long polycations.[25] In contrast, short-range attractive forces between two DNA duplexes, like depletion forces[26] and fluctuation correlation forces[27], provoke a very abrupt condensation of the DNA to a globule with solid-like density above a certain threshold concentration of crowders and/or polycations,[24,28-31] with the threshold concentration decreasing markedly with increasing concentrations of bridging nucleoid proteins.[32] However, the intermediate DNA concentrations observed in living bacteria and the gradual variations of the size of the nucleoid with environmental conditions[7-18] clearly do not support such an all-or-non mechanism.

Still another mechanism was proposed about 20 years ago on the basis of theoretical grounds.[33-35] The suggestion is that increasing amounts of non-binding spherical crowders, like nano-particles or globular neutral proteins, are able to compact the DNA gradually.[33-35]. More precisely, the overall repulsion between all components of the system leads to a segregative separation into two phases,[36] one of them being rich in DNA and poor in spherical crowders, and the other one being composed essentially of crowders and almost deprived of DNA. While the outcome of theoretical predictions depends crucially on the details of the description of the interactions amongst crowders and between crowders and the DNA,[33-35,37,38] this mechanism has recently received strong support from two series of experiments. It was indeed first shown that the addition of 5-10% (w/v) of bovine serum albumin (BSA) to the buffer compacts long DNA molecules to densities close to that of the nucleoid.[39,40] Since the surface of BSA proteins displays small positive patches and the formation of weak BSA-DNA coacervates has been reported,[41] it can admittedly not be completely excluded that the observed compaction of the DNA by BSA proteins corresponds actually to complex coacervation (associative phase separation) rather than segregative phase separation. Still, unambiguous confirmation of the efficiency of the segregative phase separation mechanism came shortly after from experiments performed with negatively charged silica nanoparticles with diameter in the range 20-135 nm, which showed that introduction of a few percents thereof in the buffer also leads to the gradual compaction of the DNA.[42]

These two series of *in vitro* experiments therefore suggest that the formation of the nucleoid *in vivo* could result from the demixing of the DNA and other globular macromolecules of the cytosol, which interact repulsively with themselves and with the DNA.[5] A survey of the molecular species found in the cytosol further indicates that this role may be played by ribosomes,[5] which are ribonucleoprotein complexes that synthesize proteins from transfer RNA (tRNA) according to templates conveyed by messenger RNA (mRNA). In their 70S functional form, ribosomes contain approximately 4500 nucleotides and 7000 amino acid residues, have a diameter of 20-25 nm, are almost uniformly negatively charged, and account for approximately 30% of the dry mass of the cell. It has furthermore long been known that most functional ribosomes are excluded from the nucleoid, a point which has been confirmed by recent in vivo microscopy experiments.[16,17,43]

Although the two series of experiments mentioned above have triggered renewed interest in this field,[44-47] much remains to be done to clarify the possible role of segregative phase separation in the compaction of the bacterial DNA inside the nucleoid. For example, the authors of the experiments with silica nanoparticles acknowledge the difficulty to estimate the effective nanoparticle volume occupancy ratio at which they observe segregation.[42] According to their calculations, this ratio lies around 15-20%, but they nevertheless do not exclude the possibility that it may instead approach the critical ratio for densely packed spheres (about 74%), because of the large uncertainty on the value of the Debye length. Recent simulations based on a coarse-grained model, where DNA is described as a freely jointed chain of blobs of radius ≈50 nm, predict that the occupancy ratio at which maximum compaction occurs is inversely proportional to the size of the crowders,[44-47] while other simulations based on a finer-grained semi-rigid model of the DNA molecule suggest instead that compaction of long DNA molecules by non-binding spherical crowders is governed by the volume occupancy ratio of the crowders and that it increases sharply up to nucleoid-like values slightly below the jamming transition.[48] This same work highlights the fact that the largest crowders demix preferentially from the DNA in systems containing crowders of different size.[48]

The purpose of the present work is to elaborate further on the predictions of this latter model[48] and to address several points, which may be deemed essential for understanding the formation of the nucleoid. First, one may wish to ascertain that the compaction of the DNA chain observed in the simulations is indeed appropriately described as a segregative phase separation.[49] According to the extension of Flory-Huggins theory to solutions containing two polymer species A and B, segregative phase separation occurs if the interaction parameter $\chi$ is positive, where $\chi$ denotes the strength of the pair interaction between A and B segments minus the average of the strengths of the pair interaction between two A segments and the pair interaction between two B segments[36,50] (segregative phase separation may actually also be driven by differences in polymer/solvent interactions,[36,51] but the coarse-grained model does not allow for such differences). If the segregative phase separation scenario is correct, significant variations of the level of compaction of the DNA chain are therefore expected upon variation of the strength of the DNA/crowder interaction compared to the strength of the DNA/DNA and crowder/crowder interactions. This is the first point addressed in the present work.

The second point deals with the geometry of the crowders. It is indeed known, that linear anionic polymers condense the DNA abruptly to a very compact globule[30,31] above the threshold concentration where depletion forces[26] overcome electrostatic repulsion between DNA duplexes, while spherical anionic nanoparticles[42] and globular anionic proteins[39,40] provoke instead a gradual compaction of the DNA to intermediate concentrations. One is therefore led to wonder how sensitive against the geometry of globular crowders the segregative phase separation mechanism may be. The present paper reports on simulations that were performed with globular crowders with different geometries (spheres, dumbbells, and octahedra), in order to get an indication thereof.

Finally, several sets of simulations were performed in order to clarify the influence of the size dispersion of the crowders on the segregative phase segregation mechanism. Indeed, it was shown in the previous work that the DNA chain and the largest crowders demix preferentially when the DNA chain interacts with crowders of different size.[48] One may wonder whether such size selectivity is responsible for the fact that functional 70S ribosomes are excluded from the nucleoid, while 30S and 50S free subunits are able to diffuse inside the DNA coil.[43]. We will report on the various simulations that were launched to get a tentative answer to this question.

## 2. Simulation models and methods

The models used in this study share several common points with those developed previously to investigate facilitated diffusion,[52-54] the interactions of H-NS proteins and DNA,[55-57], the formation of the bacterial nucleoid,[4,5,48,44-47] as well as questions dealing with spatial confinement and molecular crowding,[58,59] the collapse of DNA by combined bridging and self-adhesion,[60] and the dynamics of a DNA molecule confined into a cylindrical container and compressed by a piston.[61]

More precisely, genomic DNA is represented by a circular chain of $n = 1440$ beads of radius $a$ separated at equilibrium by a distance $l_0 = 5.0$ nm, where each bead represents 15 consecutive base pairs. The DNA chain is enclosed in a confining sphere of radius $R_0 = 120$ nm (see Fig. 1(a)), so that the concentration of nucleotides is close to the physiological value (approximately 10 mM) in spite of the 200-fold reduction in length relative to the DNA of *E. coli* cells. $N$ crowders are also enclosed in the confining sphere. For most of the simulations discussed below, crowders were taken in the form of independent spheres (see Fig. 1(b)), but several sets of simulations were run with crowders composed of two spheres (dumbbells, see Fig. 2(a)) or six spheres (octahedra, see Fig. 2(b)). The number $N$ and the size of the crowders were varied to investigate different levels of crowding, but the overall size of the crowders was usually kept in the range 15-25 nm, so as to mimic ribosomes and their free subunits.

The potential energy of the system, $E_{pot}$, is written as the sum of the internal energy of the DNA chain ($V_{DNA}$), the DNA/crowder interactions ($V_{DNA/C}$), the crowder/crowder interactions ($V_{C/C}$), the repulsive potentials that maintain the DNA chain and the crowders inside the confining sphere ($V_{wall}$), and eventually the internal energy of dumbbells and octahedral crowders ($V_C$)

$$E_{pot} = V_{DNA} + V_C + V_{DNA/C} + V_{C/C} + V_{wall} . \qquad (1)$$

$V_{DNA}$ is further expanded as the sum of 3 contributions

$$V_{DNA} = \frac{h}{2}\sum_{k=1}^{n}(l_k - l_0)^2 + \frac{g}{2}\sum_{k=1}^{n}\theta_k^2 + e_{DNA}^2 \sum_{k=1}^{n-2}\sum_{j=k+2}^{n} H(\|\mathbf{r}_k - \mathbf{r}_j\| - 2a_0) , \qquad (2)$$

where

$$H(r) = \frac{1}{4\pi\varepsilon r}\exp\left(-\frac{r}{r_D}\right), \qquad (3)$$

which describe the stretching, bending, and electrostatic energy of the DNA chain, respectively. $\mathbf{r}_k$ stands for the position of DNA bead $k$, $l_k$ for the distance between two successive beads, and $\theta_k$ for the angle formed by three successive beads. The stretching energy is aimed at avoiding a rigid rod description and has no biological meaning. $h$ was set to $1000\, k_B T / l_0^2$ to insure that $|l_k - l_0|$ remains on average of the order of $0.02\, l_0$, in spite of the forces exerted by the remainder of the system (in this work, energies are expressed in units of $k_B T$, with $T = 298\,\text{K}$). The bending rigidity constant, $g = 9.82\, k_B T$, was chosen so as to provide the correct persistence length for DNA, $\xi = g l_0 / (k_B T) \approx 49$ nm.[62] Note that $\xi$ corresponds approximately to a segment of 10 successive beads. It should also be stressed that the diameter of the confining sphere is consequently only about 5 times the persistence length of the DNA chain, so that mechanical consequences of the bending rigidity of the DNA chain may be somewhat overestimated. This is arguably the major effect of size reduction for this model. Finally, the electrostatic repulsion between DNA beads is written as a sum of Debye-Hückel potentials,[63] where $e_{DNA}$ denotes the value of the point charge placed at the centre of each DNA bead, $\varepsilon = 80\,\varepsilon_0$ is the dielectric constant of the medium, $r_D = 1.07$ nm the Debye length inside the medium (corresponding to a concentration of monovalent salts close to 100 mM, as often assumed in bacteria), and $2a_0$ the width of the (eventual) hard core of the interaction. Two different models of the DNA molecule were alternatively considered in this work. In the first model, DNA beads have a radius $a = 1.78$ nm (which was shown to lead to the correct diffusion coefficient for the DNA chain)[64] and a charge $e_{\text{DNA}} = -12.15\,\overline{e}$ (where $\overline{e}$ is the absolute charge of the electron) and the interaction potential has a soft core ($a_0 = 0$). This is the model, which was used in the previous work.[48] In the second model, DNA beads have a smaller radius $a = 1.0$ nm and a smaller charge $e_{\text{DNA}} = -7.05\,\overline{e}$, but their interaction potential has a hard core ($a_0 = a$). The diffusion coefficient of the DNA chain is consequently slightly too large for this second model, but this is of little consequence because we are essentially interested in the equilibrium properties of the model. Note also that for both models $e_{\text{DNA}}$ is significantly smaller than the net total charge carried by the phosphate groups of 15 base pairs ($-30\,\overline{e}$), which reflects the importance of counter-ion condensation.[65,66] In spite of their differences (soft core *vs* hard core, $e_{\text{DNA}} = -12.15\,\overline{e}$ *vs* $e_{\text{DNA}} = -7.05\,\overline{e}$), both models are in reasonable agreement with the Debye-Hückel approximation of the solution of the Poisson-Boltzmann equation.[63] It may also be worth emphasizing that the equilibrium separation of two DNA beads, $l_0 = 5.0$ nm, is too large compared to $r_D$ to warrant that different parts of the DNA chain will never cross. However, such crossings are rather infrequent and appear to affect the geometry of the DNA chain only to a limited extent. Finally, electrostatic interactions between nearest-neighbors are not included in eqn (2), because it is considered that they are already accounted for in the stretching and bending terms.

The internal energy of dumbbells and octahedral crowders contains only stretching contributions

$$V_C = \frac{h}{2} \sum_{K=1}^{N} \sum_{J=1}^{P-1} \sum_{M=J+1}^{P} (\|\mathbf{R}_{K,J} - \mathbf{R}_{K,M}\| - R_{J,M}^0)^2, \quad (4)$$

where $P$ denotes the number of connected spheres for each crowder ($P = 1$ for independent spheres, $P = 2$ for dumbbells, and $P = 6$ for octahedra), $\mathbf{R}_{K,J}$ the position of sphere $J$ of crowder $K$, and $R_{J,M}^0$ the equilibrium distance between spheres $J$ and $M$ of the same crowder.

In the same spirit as for DNA/DNA interactions, DNA/crowder and crowder/crowder interactions are expressed as sums of Debye-Hückel potentials

$$V_{\text{DNA/C}} = e_{\text{DNA}} \, e_{\text{C}} \sum_{k=1}^{n} \sum_{K=1}^{N} \sum_{J=1}^{P} H(\|\mathbf{r}_k - \mathbf{R}_{K,J}\| - a_0 - b_K - \delta)$$

$$V_{\text{C/C}} = e_{\text{C}}^2 \sum_{K=1}^{N-1} \sum_{J=1}^{P} \sum_{L=K+1}^{N} \sum_{M=1}^{P} H(\|\mathbf{R}_{K,J} - \mathbf{R}_{L,M}\| - b_K - b_L) , \qquad (5)$$

where $b_K$ denotes the radius of the spheres of crowder $K$, and $e_{\text{C}}$ is the electrostatic charge placed at their center. In this work, $e_{\text{C}}$ was set to $e_{\text{C}} = e_{\text{DNA}}$, as in previous work.[48] This choice, as well as the choice for the expressions of the various electrostatic potentials in eqns (2), (3), and (5), will be discussed in detail in the Results and Discussion section. Let us however emphasize right here that for $e_{\text{C}} = e_{\text{DNA}}$ the three functions $e_{\text{DNA}}^2 H(r - 2a_0)$, $e_{\text{DNA}} e_{\text{C}} H(r - a_0 - b_K - \delta)$, and $e_{\text{C}}^2 H(r - b_K - b_L)$, are equivalent to one another, except for the respective displacements $2a_0$, $a_0 + b_K + \delta$, and $b_K + b_L$. For DNA/DNA and crowder/crowder interactions, these displacements are just the sum of the hard-core radii of the interacting particles. In contrast, for DNA/crowder interactions, the displacement $a_0 + b_K + \delta$ is the sum of the hard-core radii of the interacting particles only in the limit where $\delta = 0$, which corresponds to the 'symmetric' case studied previously.[48] This symmetric case $\delta = 0$ is characterized by the fact that the repulsion potential between a DNA bead and a crowding sphere is the median of the repulsion potential between two DNA beads and the repulsion potential between two crowding spheres. On the other hand, the symmetry of the interactions is broken towards comparatively more repulsive (respectively, less repulsive) DNA/crowder interactions for $\delta > 0$ (respectively, $\delta < 0$). The dissymmetry coefficient $\delta$ will play a central role in the discussions of section 3.

Finally, $V_{\text{wall}}$ is written in the form

$$V_{\text{wall}} = \zeta \left( \sum_{k=1}^{n} f(\|\mathbf{r}_k\|) + \sum_{K=1}^{N} \sum_{J=1}^{P} f(\|\mathbf{R}_{K,J}\|) \right) , \qquad (6)$$

where the repulsive force constant $\zeta$ was set to $1000 k_B T$ and the function $f(r)$ is defined according to

if $r \leq R_0$ : $f(r) = 0$

if $r > R_0$ : $f(r) = \left( \dfrac{r}{R_0} \right)^6 - 1.$ \qquad (7)

The dynamics of the system was investigated by integrating numerically overdamped Langevin equations. Practically, the updated positions at time step $n+1$ were computed from the positions at time step $n$ according to

$$\mathbf{r}_k^{(n+1)} = \mathbf{r}_k^{(n)} + \frac{\Delta t}{6\pi\eta a}\mathbf{f}_k^{(n)} + \sqrt{\frac{2\,k_B T\,\Delta t}{6\pi\eta a}}\,x_k^{(n)}$$

$$\mathbf{R}_{K,J}^{(n+1)} = \mathbf{R}_{K,J}^{(n)} + \frac{\Delta t}{6\pi\eta b_K}\mathbf{F}_{K,J}^{(n)} + \sqrt{\frac{2\,k_B T\,\Delta t}{6\pi\eta b_K}}\,X_K^{(n)},$$

(8)

where $\Delta t = 20\,\mathrm{ps}$ is the integration time step, $\mathbf{f}_k^{(n)}$ and $\mathbf{F}_{K,J}^{(n)}$ are vectors of inter-particle forces arising from the potential energy $E_{\mathrm{pot}}$, $T = 298\,\mathrm{K}$ is the temperature of the system, $x_k^{(n)}$ and $X_K^{(n)}$ are vectors of random numbers extracted from a Gaussian distribution of mean 0 and variance 1, and $\eta = 0.00089$ Pa s is the viscosity of the buffer at 298 K.

After each integration step, the position of the centre of the confining sphere was adjusted slightly, so as to coincide with the centre of mass of the DNA molecule. It was indeed observed that without this centering step the DNA coil sometimes sticks for long times to the confining sphere, which alters significantly its degree of compaction and the computed density profiles. Centering was therefore introduced in the simulation scheme, in order to prevent this possibility and ensure more meaningful comparisons between different runs. We note in passing that the location of the nucleoid close to (or away from) the membrane is in itself an interesting but complex question. Indeed, standard arguments predict that if both the compacted DNA and crowding macromolecules could be considered as solid particles, then the (comparatively larger) compacted DNA should have a marked tendency to localize close to the membrane, in order to increase the space available for the (comparatively smaller) crowding macromolecules,[26] which obviously contradicts the experimental fact that the nucleoid is most frequently observed away from the membrane. The ability of crowding macromolecules to penetrate inside the DNA coil thus likely plays a role in the positioning of the nucleoid inside the cell, as consequently also does the degree of compaction of the DNA molecule. Work in this direction is in progress, but in the present work it was chosen for clarity to disentangle DNA compaction and DNA location through the centering scheme.

Each point shown in the figures discussed in section 3 was obtained from a single trajectory, which was integrated for times as long as 100 ms close to the jamming threshold. The mean radius of gyration of the DNA chain, $R_g$, as well as the mean density profiles for each species, were computed after equilibration of each conformation. It is estimated that the uncertainty for the computed values of the mean radius of gyration $R_g$ is of the order of ±1 nm away from the inflexion point of the curves and ±3 nm closer to the inflexion point, where larger oscillations are observed. The estimated uncertainty for enrichment coefficients $Q_{X/Y}$ (subsection 3.3) is of the order of ±5%.

It is important to note that the model for DNA described above is significantly finer-grained compared to the model proposed in Refs. [44-47]. Indeed, each bead represents here 15 consecutive base pairs of the DNA molecule, so that the persistence length of DNA (≈50 nm) is equivalent to 10 beads and the bending rigidity has to be taken into account. In contrast, in Refs. [44-47] each DNA bead represents a blob of radius at least 50 nm (meant to consist of supercoiled DNA strands and DNA-bound proteins) and bending rigidity can be neglected in first approximation. This difference in the coarse-graining of the DNA molecule has an important consequence when crowders are assumed to consist of molecular complexes of radius ≈10 nm, like ribosomes. Indeed, spheres representing the crowders are consequently larger than DNA beads in the present model, while they are significantly smaller than DNA blobs in the model of Refs. [44-47]. The crucial point is that simulations performed with this latter model led to rather different results, depending on whether the size of crowders was assumed to be smaller (the "bacterial chromosome limit") or larger (the "protein folding limit") than the size of the connected beads.[45,46] It can consequently be expected that the results obtained with the present model will, somewhat paradoxically, be closer to those pertaining to the "protein folding limit" of Refs. [45,46] rather than the "bacterial chromosome limit". This point will be discussed further throughout the remainder of this paper.

**3.     Results and discussion**

**3.1    Sensitivity of DNA compaction against dissymmetry of repulsive interactions**

Simulations discussed in the present paper consisted in (i) letting the DNA chain equilibrate inside the confining sphere, (ii) introducing the crowders at random non-overlapping positions inside the sphere, and (iii) letting the system equilibrate again, which eventually resulted in compaction of the DNA chain. This is illustrated in Fig. 1, which shows a representative conformation of the DNA chain after equilibration inside the confining sphere (Fig. 1(a)), the system after introduction of 1830 spherical crowders inside the confining sphere containing the equilibrated DNA chain (Fig. 1(b)), and a representative snapshot of the conformation of the DNA chain after equilibration of the full system (Fig.

1(c)). The degree of compaction of the DNA chain was quantified by the mean radius of gyration of the chain, $R_g$. As indicated in Fig. 1, $R_g$ is of the order of 83 nm in the absence of crowders and decreases down to about 64 nm after equilibration with the set of crowders shown in Fig. 1(b) and to about 50 nm when 500 additional small crowders are introduced in the confining sphere (Fig. 1(d)).

It was shown previously[48] that for spherical crowders with homogeneous radius $b_K = b$ the degree of DNA compaction is actually governed by the volume fraction of the crowders, $\rho$, computed according to

$$\rho = \frac{1}{R_0^3} \sum_{K=1}^{N} (b_K + \Delta b)^3 \,, \tag{9}$$

where $b_K + \Delta b$ denotes the effective radius of the crowders, that is, half the distance where the repulsion between two crowders is equal to the thermal energy $k_B T$ ($\Delta b = 1.8$ nm for $e_C = -12.15\,\overline{e}$, $\Delta b = 1.3$ nm for $e_C = -7.05\,\overline{e}$). One of the interesting points in describing crowder/crowder interactions through the potential in eqn (5), instead of the more usual DLVO potential[67,68]

$$W^{\text{DLVO}}(r) = \frac{e_M^2}{4\pi\varepsilon r(1+\frac{b_K}{r_D})^2} \exp(-\frac{r-2b_K}{r_D}) \,, \tag{10}$$

where $e_M$ is the total charge of the sphere, is precisely that $\Delta b$ does not depend on $b_K$ for the potential in eqn (5), while it does for the DLVO potential. By running simulations with different values of $N$ and $b$, it could therefore be shown clearly that $R_g$ decreases almost linearly with $\rho$ down to $R_g \approx 64$ nm for $\rho \approx 0.55$, before dropping sharply to $R_g \approx 50-55$ nm for $\rho \approx 0.65$, just below the jamming threshold at $\rho \approx 0.75$.[48] As anticipated in the Simulation Models and Methods section, it is interesting to note that this result is indeed closer to the "protein folding limit" of Refs. [45,46], for which compaction was also shown to be governed by the volume fraction of the crowders, rather than the "bacterial chromosome limit", for which it is the ratio of the volume fraction and the size of the crowders that matters.[45,46] It is nonetheless emphasized that the size of the DNA coil evolves much more smoothly and gradually with the volume fraction of the crowders for the "protein folding limit" of Refs. [45,46] than for the present model, a difference which is probably ascribable to the bending rigidity, which in the present model opposes compaction efficiently up to the jamming threshold, where intermolecular interactions finally become predominant.

In the present work, we will take advantage of another property of the potential functions in eqns (2), (3) and (5), namely that the parameter $\delta$ in eqn (5) correlates tightly with the interaction parameter $\chi$ of Flory-Huggins solution theory.[36,50] Indeed, as pointed out in the Simulation Models and Methods section, the symmetric case $\delta = 0$ is characterized by the fact that the repulsion potential between a DNA bead and a crowding sphere is the median of the repulsion potential between two DNA beads and the repulsion potential between two crowding spheres, while the symmetry of the interactions is broken towards comparatively more repulsive (respectively, less repulsive) DNA/crowder interactions for $\delta > 0$ (respectively, $\delta < 0$). Increasing (respectively, decreasing) $\delta$ therefore leads to an increase (respectively, a decrease) of $\chi$. In contrast, variations of $\delta$ do not affect the volume ratio of crowders, $\rho$, which was shown to govern the compaction of the DNA chain for $\delta = 0$.[48] According to solution theory, demixing occurs only for positive values of the interaction parameter $\chi$.[36,50] If the description of the compaction of the DNA chain as a segregative phase separation is appropriate,[48] then the symmetric case $\delta = 0$ corresponds to positive values of $\chi$ for crowder densities close to the jamming threshold. Moreover, solution theory predicts that the extent of demixing between two solutes evolves continuously with the value of $\chi$.[36,50] Increased (respectively, decreased) compaction of the DNA chain is therefore expected for positive (respectively, negative) values of $\delta$.

We accordingly performed simulations with different values of $\delta$ and $\rho$ to check the appropriateness of the description of the compaction of the DNA chain in terms of segregative phase separation and DNA/crowders demixing. These simulations were run with the soft core model for the DNA chain ($a = 1.78$ nm, $e_{DNA} = -12.15 \overline{e}$, $a_0 = 0$, and $\Delta b = 1.8$ nm) and with $N = 500$ spherical crowders having the same radius $b$. Four different values of $b$ were plugged in the simulations, namely $b = 9.0$, 10.0, 11.0, and 11.5 nm, which correspond to crowder volume ratios $\rho = 0.36$, 0.48, 0.61, and 0.68, respectively and for each value of $b$, simulations were run for values of $\delta$ ranging from -1.5 nm to 2.0 nm. The plots of $R_g$ vs $\delta$ are shown in Fig. 3 for the four different values of $b$. It is seen in this figure that the level of compaction of the DNA chain indeed increases with $\delta$, as expected for the segregative phase separation scenario. For $b = 11.5$ nm, that is, very close to the jamming transition, the evolution of $R_g$ vs $\delta$ is almost step-like, with $R_g$ decreasing by about 50% (from $\approx 75$ nm down to $\approx 40$ nm) upon increase of $\delta$ from -0.5 nm to 0.5 nm. The evolution of $R_g$ with $\delta$ is smoother further away from the jamming transition but remains significant even at rather moderate crowder volume ratios. For example, it is seen in Fig. 3 that for $\rho = 0.36$ (about half the value of $\rho$ at the jamming transition), a dissymmetry coefficient $\delta = 1.0$ nm leads to a radius of gyration of the DNA chain as small as $R_g \approx 55$ nm, which is the value observed close to the jamming transition for the symmetric case $\delta = 0$. For the sake of completeness, we recall here that results obtained for different crowder size $b$ and crowder number $N$ ($500 \leq N \leq 3000$) were shown to superpose when plotted as a function of volume fraction $\rho$.[48] Finite size effects related to the relatively modest number of crowders used in the simulations discussed above ($N = 500$) therefore probably do not affect significantly the results.

In conclusion, Fig. 3 supports the description of the compaction of the DNA chain as a segregative phase separation mechanism. For symmetric repulsive interactions ($\delta = 0$), the demixing of DNA beads and crowding spheres is attributable to the connectivity of DNA beads, because no compaction is ever observed when the bonds between DNA beads are broken (result not shown). Fig. 3 however indicates that the level of DNA compaction is also very sensitive to the dissymmetry of repulsive interactions. In particular, strong compaction can be obtained far from the jamming threshold, provided that the symmetry of repulsive interactions is sufficiently displaced towards stronger DNA/crowder repulsion.

### 3.2  Sensitivity of DNA compaction against the shape of crowders

As mentioned in the Introduction, the shape of anionic crowders affects profoundly their ability to compact the DNA macromolecule. Indeed, linear polymers condense DNA abruptly to a very compact globule above a certain concentration threshold,[30,31] while spherical nanoparticles[42] and globular proteins[39,40] provoke instead a gradual and softer compaction. In order to get some insight into the sensitivity of the compaction properties of anionic crowders against their shape, we performed several sets of simulations with non-spherical crowders. More precisely, we considered dumbbells, which are composed of two spheres of radius $b$ separated at equilibrium by a distance $R^0_{1,2} = b$, and octahedral crowders, which are composed of six spheres of radius $b$ separated at equilibrium by a distance $R^0_{J,M} = \sqrt{2}b$ from their neighbors and $R^0_{J,M} = 2b$ from opposite spheres. These simulations were run with the hard core model for the DNA chain ($a = 1.0$ nm, $e_{\text{DNA}} = -7.05\,\bar{e}$, $a_0 = a$, and $\Delta b = 1.3$ nm) and with $N = 500$ crowders. Zooms on representative conformations of the systems are shown in Fig. 2.

Four different values of $b$ were plugged in the simulations performed with dumbbells, namely $b = 8.2$, 9.1, 10.0, and 10.3 nm, which correspond to crowder volume ratios $\rho = 0.40$, 0.53, 0.68, and 0.73, respectively (the volume of the intersection of the two spheres of effective radius $b + \Delta b$ is obviously counted only once in the calculation of $\rho$). For each value of $b$, simulations were run for values of $\delta$ ranging from 0 to 1 nm and the mean radius of gyration of the DNA chain was computed after equilibration of each system. The plots of $R_g$ vs $\delta$ are shown in Fig. 4 for the four different values of $b$. It is seen in this figure, that dumbbell crowders share two important properties with spherical crowders, namely that the level of compaction of the DNA chain increases with $\delta$ and the evolution of $R_g$ vs $\delta$ is sharper closer to the jamming transition. There is, however, one important difference between Figs. 3 and 4. Indeed, for values of $b$ close to the jamming transition, the inflexion point of the $R_g$ vs $\delta$ curves is located close to $\delta = 0$ (symmetric repulsive potentials) for spherical crowders, while it is shifted to $\delta \approx 0.35$ for dumbbells. This implies that the symmetry of repulsive interactions must be displaced towards stronger DNA/crowder repulsion for significant compaction of DNA to take place. We will come back to this point shortly.

Four different values of $b$ were also plugged in the simulations performed with octahedral crowders, namely $b = 5.0$, 5.5, 6.2, and 6.7 nm, which correspond to volume ratios $\rho = 0.28$, 0.36, 0.49, and 0.60, respectively, and for each value of $b$ simulations were run for values of $\delta$ ranging from 0 to 2 nm. The corresponding plots of $R_g$ vs $\delta$ are shown in Fig. 5. It is seen in this figure that the compaction of the DNA chain increases with $\delta$ and the evolution of $R_g$ vs $\delta$ is sharper closer to the jamming transition, as for spherical crowders and dumbbells. The inflexion point of the $R_g$ vs $\delta$ curves is moreover also shifted towards positive values of $\delta$, with the shift being significantly larger for octahedral crowders ($\delta \approx 1.25$) than for dumbbells ($\delta \approx 0.30$). For the sake of an easier comparison, representative plots of $R_g$ vs $\delta$ for spherical, dumbbell, and octahedral crowders are superposed in Fig. 6 for both heavy and light crowding conditions. The mere displacement of the curves towards larger values of $\delta$ when going from spherical to octahedral through dumbbell crowders is clearly seen in this figure. These simulations therefore raise the question, why compaction of the DNA chain by dumbbells and octahedral crowders requires the displacement of the symmetry of repulsive interactions towards stronger repulsion between DNA beads and individual crowding spheres.

Most probably, the answer to this question has to be sought in the fact that not only the shape, but also the distribution of charges, is different for spherical, octahedral, and dumbbell crowders (remember that a charge $e_C = e_{DNA}$ is placed at the center of each crowding sphere). This hypothesis can be tested quite straightforwardly in the case of octahedral crowders, because the symmetry of an octahedron is not far from the symmetry of a sphere. Instead of comparing the repulsive potentials between DNA beads and individual crowding spheres, as in subsection 3.1 above and Fig. 3 of the previous work,[48] one can therefore consider the evolution, as a function of the distance $r$ between their centers, of the repulsion energy between a DNA bead and a full octahedron, or between two full octahedra. For the sake of simplicity, the geometry of each octahedron is frozen to its equilibrium conformation and the repulsion energy is minimized with respect to all orientations of the octahedra at fixed $r$. The result for $b = 6.7$ nm and $\delta = 0$ is shown in Fig. 7, where the blue long-dashed curve represents the repulsion energy between two DNA beads, the green short-dashed curve the repulsion energy between a DNA bead and an octahedron, and the red solid curve the repulsion energy between two octahedra. When calculated as sketched above, the interaction energy between a DNA bead and an octahedron is actually close to the repulsive potential between a DNA bead and a single crowding sphere of radius $3b/2$, as can be checked in Fig. 7, where the green short-dashed curve almost superposes on a grey one, which represents the repulsion energy between a DNA bead of radius $a = 1.0$ nm and a crowding sphere of radius $b = 9.75$ nm. Note that we did not seek for a formal derivation of this empirical result, which holds for all investigated values of $b$. Similarly, the repulsion energy between two octahedra composed of spheres of radius $b = 6.7$ nm is very close to the repulsive potential between two crowding spheres of radius $b = 10.75$ nm, as can be checked in Fig. 7, where this latter potential is represented by a grey solid curve, which nearly superposes on the red one. The curve representing the repulsion energy between a DNA bead and an octahedron is therefore clearly shifted towards lower values of $r$ compared to the median of the curves representing the repulsion energy between two DNA beads and the repulsion energy between two octahedra. Stated in other words, the potentials that describe the interactions between DNA beads and full octahedra are not symmetric for $\delta = 0$, that is for symmetric interactions between DNA beads and individual crowding spheres. A given amount of dissymmetry must instead be introduced in the interactions between DNA beads and individual crowding spheres ($\delta > 0$) to restore the symmetry of the potentials involving DNA beads and full octahedra. For $b = 6.7$ nm, the corresponding value of the dissymmetry coefficient is $\delta = 10.75 - 9.75 = 1.0$ nm, which is indeed of the correct order of magnitude compared to the location of the inflexion points in Fig. 5 ($\delta \approx 1.25$).

Similarly, when plotted as a function of the distance between their centers of mass, the interaction energy (computed as described above) between a DNA bead and a dumbbell composed of two spheres of radius $b = 10.0$ nm is very close to the repulsive potential between a DNA bead and a single crowding sphere of radius $b = 9.5$ nm, while the interaction energy between two dumbbells is very close to the repulsive potential between two crowding spheres of radius $b = 10.0$ nm. For dumbbells composed of two spheres of radius $b = 10.0$ nm, the value of the dissymmetry coefficient, which is required to get symmetric interactions between DNA beads and dumbbell crowders, is therefore $\delta = 10.0 - 9.5 = 0.5$ nm. This is again of the correct order of magnitude compared to the location of the inflexion points in Fig. 4 ($\delta \approx 0.30$).

In conclusion, simulations performed with globular but non-spherical crowders (dumbbells and octahedra) provide evidence that the compaction of the DNA chain by these crowders is also driven by the symmetry/dissymmetry of the potentials describing the repulsive interactions between DNA beads and the crowders. These simulations consequently also support the description of the compaction of the DNA chain as a segregative phase separation and suggest that the compaction mechanism is not too sensitive to the precise shape of the crowders.

### 3.3 Influence of crowders' size dispersion on DNA/crowders demixing

As mentioned in the Introduction, it was shown previously that the DNA and the largest crowders demix preferentially when the DNA chain interacts with crowders of different size[48], thus raising the question whether this may explain the experimental observation that functional 70S ribosomes are excluded from the nucleoid, while 30S and 50S free subunits are able to diffuse inside the DNA coil.[43] (note that the influence of crowders' size dispersion was not investigated for the "protein folding limit" of Refs. [45,46], while the "bacterial chromosome limit" displays only limited sensitivity against size dispersion[44]). Several sets of simulations were launched to answer this question, which all involved the soft core model for the DNA chain ($a = 1.78$ nm, $e_{DNA} = -12.15\,\overline{e}$, $a_0 = 0$, and $\Delta b = 1.8$ nm), spherical crowders, and symmetric repulsive interactions ($\delta = 0$).

In the first set of simulations, the total number of crowders was set to $N = 2000$ and their volume occupancy ratio to $\rho = 0.66$, as if all crowders were of radius $b = 6.5$ nm. This value of $\rho$ is only slightly smaller than the jamming threshold for solid spheres and leads to strong compaction of the DNA chain.[48] Crowders were however divided into a first set of $N_B = 1000$ big spheres of radius $b_B$ and a second set of $N_S = 1000$ smaller spheres of radius $b_S$. $b_B$ was varied from 6.7 to 8.1 nm with increments of 0.2 nm and, for each value of $b_B$, the value of $b_S$ was adjusted according to eqn (9) to match the volume ratio $\rho = 0.66$. At time $t = 0$, the crowders were placed at random non-overlapping positions inside the confining sphere containing the relaxed DNA chain, so that all species (DNA beads, big and small crowding spheres) initially had rather similar and nearly uniform density distributions $p_X(r)$ (defined so that the mean number of particles of species $X$ with center located in the distance interval $[r, r+dr]$ from the center of the confining sphere is $4\pi n_X p_X(r) r^2 dr$, where $n_X = n$, $N_B$, or $N_S$, denotes the total number of particles of type $X$). Upon relaxation of the full system, the DNA chain compacts progressively in the central region of the confining sphere, while a certain number of crowding spheres move simultaneously towards its periphery. This is illustrated in the inset of Fig. 8, which shows the resulting mean density distributions for the equilibrated system with $b_B = 7.1$ nm and $b_S = 5.77$ nm. In particular, it is seen in this plot that the density distribution of big crowding spheres is smaller than that of small spheres in the central region of the confining sphere, meaning that big crowders are expelled preferentially from the DNA coil during its compaction. This differential effect can be quantified by computing $Q_{S/B}$, the enrichment inside the DNA coil of small crowding spheres relative to big ones

$$Q_{S/B} = \frac{N_S(r < r_{thresh})}{N_B(r < r_{thresh})} \frac{N_B}{N_S} \,. \tag{11}$$

In this expression, $N_B(r < r_{thresh})$ and $N_S(r < r_{thresh})$ denote the number of big and small crowding spheres with center located at a distance smaller than $r_{thresh}$ from the center of the confining sphere. In the following, we will use $r_{thresh} = R_0 / 2 = 60$ nm, which is indicated as a dot-dashed vertical line in the insert of Fig. 8. The evolution of $Q_{S/B}$ as afunction of $b_B / b_S$ is shown in the main plot of Fig. 8. It is seen in this plot that the enrichment inside the DNA coil of small crowding spheres relative to big ones increases almost linearly with $b_B/b_S$ over a relatively narrow range, before saturating for values of $b_B/b_S$ close to 2 and decreasing slowly above this value. The increase of $Q_{S/B}$ for values of $b_B/b_S$ slightly larger than 1 is a clear indication that the interaction parameter $\chi$ of DNA and the crowders increases with the size of the crowders, so that bigger crowders demix preferentially from the DNA coil. The saturation around $Q_{S/B} \approx 2$ for $b_B/b_S \geq 2$ may in turn be interpreted as a kinetic effect. Indeed, because of their size, big crowders are much less mobile than smaller ones in systems close to the jamming threshold. If the difference in radii is sufficiently large, smaller crowders consequently move outside from the compacting DNA coil faster than big ones, even if their interaction coefficient is smaller, and a certain number of big crowders remain eventually trapped inside the DNA coil before the system is able to equilibrate thermodynamically.

In the second set of simulations, the radii of big and small crowding spheres were set to $b_B = 10.0$ nm and $b_S = 5.0$ nm, respectively, and the number of big crowding spheres to $N_B = 400$, while the number of small crowding spheres, $N_S$, was varied from 1300 to 2100. The evolution, as a function of $N_S$, of the mean radius of gyration of the DNA chain, $R_g$, and the enrichment inside the DNA coil of small crowders relative to big ones, $Q_{S/B}$, is displayed in Fig. 9. As in Figs. 3, 4 and 5, $R_g$ drops again sharply close to the jamming threshold. Quite interestingly, the drop of $R_g$ is here accompanied by a similar drop of $Q_{S/B}$. At moderate DNA compaction ($N_S = 1500$, $\rho = 0.65$, and $R_g \approx 64$ nm), the enrichment inside the DNA coil of small crowders relative to big ones is indeed close to 10, meaning that big crowders are almost completely excluded from the DNA coil. However, at stronger DNA compaction ($N_S = 2000$, $\rho = 0.74$, and $R_g \approx 52$ nm), $Q_{S/B}$ drops down to almost 1, which indicates very little enrichment. This second set of simulations therefore confirms that when crowders of different size are present in the confining sphere, the final density distribution of each species results from a competition between thermodynamic forces, which tend to let bigger crowders escape the compacting DNA coil preferentially, and steric hindrance, which slows down the motion of big crowders relative to smaller ones. Very close to the jamming threshold, the motion of all crowders, whether big or small, is strongly hindered, so that it is essentially the thin DNA chain that moves in a quasi-static network of crowders to achieve compaction, thus explaining why very little enrichment is observed in this limit.

Finally, we performed a third set of simulations, which may reflect more accurately the actual content of the cytoplasm of bacteria than the first two sets. Three different crowder species were taken into account in these latter simulations, namely $N_B = 320$ big crowders of radius $b_B = 10.0$ nm, which represent functional ribosomes, $N_M = 160$ medium-sized crowders of radius $b_M = 7.94$ nm, which represent free ribosomal subunits, and a variable number $1250 \leq N_S \leq 2050$ of small crowders of radius $b_S = 5.0$ nm, which represent the other macromolecules present in the cytosol. A typical initial conformation of the system with $N_S = 1350$ small crowders is shown in vignette (b) of Fig. 1 and representative conformations of the DNA chain after relaxation of the full system in vignettes (c) ($N_S = 1350$) and (d) ($N_S = 1850$) of the same figure. Also shown in Fig. 10 are the evolution, as a function of $N_S$, of the mean radius of gyration of the DNA chain, $R_g$, the enrichment inside the DNA coil of small crowders relative to big and medium-sized ones,

$$Q_{S/(M+B)} = \frac{N_S(r < r_{thresh})}{N_M(r < r_{thresh}) + N_B(r < r_{thresh})} \frac{N_M + N_B}{N_S} , \qquad (12)$$

and the enrichment of medium crowders relative to big ones,

$$Q_{M/B} = \frac{N_M(r < r_{thresh})}{N_B(r < r_{thresh})} \frac{N_B}{N_M} . \qquad (13)$$

It is seen in this figure that the evolution of $Q_{S/(M+B)}$ is similar to the evolution of $Q_{S/B}$ in the previous set of simulations, in the sense that the drop of $Q_{S/(M+B)}$ close to the jamming threshold parallels the drop of $R_g$. The evolution of $Q_{M/B}$ is instead somewhat different, although $Q_{M/B}$ is also very large (of the order of 10) at moderate DNA compaction ($N_S = 1250$, $\rho = 0.62$, and $R_g \approx 66$ nm), meaning that big crowders are almost completely and quite selectively excluded from the DNA coil under these conditions. However, $Q_{M/B}$ drops sharply down to about 1, and size selectivity among big and medium-sized crowders is lost, at substantially lower values of $N_S$ compared to $Q_{S/(M+B)}$ ($N_S \approx 1550$, $\rho = 0.67$, and $R_g \approx 62$ nm, against $N_S \approx 2050$, $\rho = 0.76$, and $R_g \approx 52$ nm). This indicates that, in the presence of a large number of small crowders, the density distributions of the remaining crowders inside the compacted DNA coil decrease with their relative sizes only at rather moderate total crowder concentrations, where all crowders are able to diffuse almost freely. However, as soon as steric hindrance comes into play, kinetic and caging effects oppose thermodynamic forces and eventually overwhelm them.

In conclusion, simulations performed with spherical crowders of different radii show that the biggest crowders are expelled selectively from the DNA coil only at relatively low total crowder concentrations, that is, for conditions that favor only moderate DNA compaction. In contrast, the size of the nucleoid of living bacteria agrees with the strong compaction of the DNA chain predicted by the model at large crowder concentrations, close to the jamming threshold. The hypothesis that the bacterial cytoplasm is close to the jamming threshold is further supported by the recent observation that the motion of macromolecules is much slower in the bacterial cytoplasm than in water and in eukaryotic cells[69] and exhibits non-Gaussian sub-diffusive behavior[70], as well as the observation that the cytoplasm itself displays properties that are characteristic of glass-forming liquids.[71] On the basis of the simulations, and owing to the probable proximity to the jamming threshold, one consequently expects that the expulsion of large crowders from the nucleoid is only mildly sensitive to their size. This suggests in turn that size effects cannot be responsible for the fact that functional 70S ribosomes are expelled from the nucleoid, while 30S and 50S free subunits diffuse inside the DNA coil.[43]. Moreover, simulations show that it is essentially the volume occupancy ratio of crowders that matters for the purpose of DNA compaction, not their exact size, so that similar compaction ratios are expected when ribosomes are in their 70S functional form or separated into free 30S and 50S subunits. This implies that, in the absence of full ribosomes, free ribosomal subunits should be expelled from the nucleoid while compacting the DNA coil. This prediction is, however, in contradiction with recent experiments involving cells treated with rifampicin (an antibiotic that causes all ribosomes to convert to free 30S and 50S subunits), which showed fully decondensed nucleoids extending throughout the cell.[9-11,14,16-18] One is thus led to the conclusion that the interaction parameter $\chi$ of DNA and free subunits is probably either zero or negative, meaning that free subunits tend to associate with DNA rather than segregate, while that of DNA and functional ribosomes is positive, meaning that DNA and full ribosomes tend to segregate. This can happen, for example, if the faces of the two subunits, which bind together to form a functional ribosome, are also able to bind DNA duplexes weakly and unspecifically, while the rest of their surfaces is not.

**4.     Conclusions**

The work reported in this paper elaborates on the conjecture that the formation of the bacterial nucleoid may result from a segregative phase separation mechanism driven by the demixing of the DNA coil and non-binding globular macromolecules present in the cytoplasm, presumably functional ribosomes. Liquid-liquid phase separation is being increasingly recognized as one of the important organizers of the cytoplasm,[72-74] but most examples up to date, like the formation of the nucleolus, the centrosomes, and stress granules, actually involve associative phase separation (complex coacervation), where different components selectively attract each other and form regions enriched in these components (droplets) surrounded by the remaining species-poor cytoplasm. For the bacterial nucleoid, we argue here that it is instead the overall repulsion between the components, which creates a phase rich in DNA and poor in the other macromolecule (the nucleoid) and a second phase almost deprived of DNA but with large concentrations of the other macromolecule (the rest of the cytosol). Note that associative and segregative phase separations share the common property that the resulting phases are able to exchange many molecular species very rapidly, in sharp contrast with membrane bound organelles.

Let us first mention that the results discussed here are in better agreement with the "protein folding limit" than the "bacterial chromosome limit" of a model proposed previously.[44-47] The reason is that the "protein folding limit" of Refs. [45,46] was defined as the case where crowders are larger than the hard spheres composing the polymer chain, which is also the case for the model proposed in this work, while the "bacterial chromosome limit" corresponds to the opposite case where crowders are smaller than the hard spheres composing the polymer chain. The term "bacterial chromosome limit" was introduced in Refs [44-47] because it was considered that each hard sphere of the polymer chain represents a blob of radius larger than 50 nm consisting of supercoiled DNA strands and DNA-bound proteins, while most other macromolecular complexes have a smaller size. The implicit assumption underlying the "bacterial chromosome limit" of the model of Refs. [44-47] is consequently that blobs of radius 50 nm (or more) are mostly incompressible. It turns out that the finer-grained model proposed here suggests that this is not the case and that DNA blobs with a typical size of a few tens of nm can actually be compacted to an important extent. If this is correct, then the "bacterial chromosome limit" of the model of Refs. [44-47] is just too coarse-grained to describe adequately the bacterial chromosome, while the "protein folding limit" is a more reasonable approximation, provided that one considers, as is done in the present work, that each sphere of the polymer represents a short track of the DNA duplex (note that it is then important to take the bending rigidity of the DNA molecule into account, as in the present model).

The simulations discussed in the main body of this paper display a high sensitivity against the dissymmetry of DNA/DNA, DNA/crowder, and crowder/crowder repulsive interactions, thereby supporting the description of the compaction of the DNA chain as a phase separation mechanism, especially as simulations performed with dumbbells and octahedra back up the results obtained with spherical crowders. However, these findings also imply that a definitive confirmation of the segregative phase separation scenario of bacterial nucleoid formation will probably have to await a thorough examination of the interactions between actual macromolecules *in vivo*, which probably represents a rather difficult challenge in the highly charged electrolyte formed by the cytosol. In contrast, the prediction of larger compaction and higher sensitivity against external factors closer to the jamming threshold is perhaps easier to bring out experimentally.

Moreover, simulations performed with crowders of different sizes suggest that the final density distribution of each species results from the competition between thermodynamic forces, which tend to let bigger crowders escape the compacting DNA coil preferentially, and steric hindrance, which slows down the motion of big crowders relative to smaller ones. As a consequence, the model predicts that bigger crowders are expelled selectively from the nucleoid only at rather moderate total crowder concentrations. This prediction may perhaps not be too difficult to check experimentally, for example by performing *in vitro* experiments with anionic nanoparticles of different sizes. Simulations furthermore suggest that the interaction parameter $\chi$ of DNA and free ribosomal subunits is either zero or negative, while that of DNA and functional ribosomes is positive, a point which may eventually receive independent confirmation.

Last but not least, let us mention that it is quite possible that several mechanisms actually work together to compact the bacterial nucleoid and that the segregative phase separation scenario discussed here represents only the first level of compaction, which affects uniformly the whole genome, and on top of which more specialized mechanisms eventually work. In particular, it is believed that the nucleoid of *E. coli* cells is divided into four different regions, called macro-domains, with the property that contacts between DNA sites belonging to the same domain are much more frequent than between DNA sites belonging to different domains.[75-78] A certain number of nucleoid proteins are responsible for the organization of each of these domains and modulate their physical properties quite sensitively.[79-81] For example, the MatP protein is responsible for the organization of the Ter domain, which contains the replication terminus.[79,80] In the absence of MatP, the DNA in the Ter domain is less compacted, has larger mobility, and segregates earlier in the cell cycle.[79,80] Such considerations suggest a multilayered formation of the nucleoid, with segregative phase separation inducing a general but partial compaction of the DNA coil and more specific mechanisms being responsible for the finer organization and additional compaction.

**Conflicts of interest**

There are no conflicts of interest to declare.

# FIGURE CAPTIONS

**Figure 1** : Representative snapshots of simulations performed with spherical crowders. **(a)** Equilibrated conformation of the DNA chain (red beads) inside the confining sphere before introduction of the crowders. Only a quarter of the confining sphere is shown, as in vignettes (c) and (d). **(b)** Initial conformation of the system after introduction of 320 crowders with radius 10.0 nm (green spheres), 160 crowders with radius 7.94 nm (yellow spheres), and 1350 crowders with radius 5.0 nm (cyan spheres), at random non-overlapping positions inside the confining sphere containing the equilibrated DNA chain. The confining sphere is not shown. **(c)** Conformation of the DNA chain after equilibration of the system shown in vignette (b). Crowders are not shown. **(d)** Conformation of the DNA chain after equilibration of a system similar to that shown in vignette (b), except that the number of crowders with radius 5.0 nm is 1850 instead of 1350. Crowders are not shown.

**Figure 2** : Zooms on representative snapshots of simulations performed with $N=500$ dumbbells **(a)** and $N=500$ octahedral crowders **(b)**. Each vignette shows a 100 nm × 100 nm section. Dumbbells are composed of two spheres of radius $b=10$ nm and octahedral crowders of 6 spheres of radius $b=5.5$ nm. The small red beads represent the DNA chain. The crowders are colored randomly for the sake of clarity.

**Figure 3** : Plot of $R_g$, the mean radius of gyration of the DNA chain, as a function of $\delta$, the dissymmetry of the repulsive electrostatic potential (eqn (5)), for different values of $b$, the radius of spherical crowders. Simulations were run with the soft core model for the DNA chain and $N=500$ spherical crowders having the same radius $b$. $R_g$ was computed after equilibration of the full system.

**Figure 4** : Plot of $R_g$, the mean radius of gyration of the DNA chain, as a function of $\delta$, the dissymmetry of the repulsive electrostatic potential (eqn (5)), for different values of $b$, the radius of crowding spheres. Simulations were run with the hard core model for the DNA chain and $N=500$ dumbbells, each dumbbell being composed of two spheres of radius $b$. $R_g$ was computed after equilibration of the full system.

**Figure 5**: Plot of $R_g$, the mean radius of gyration of the DNA chain, as a function of $\delta$, the dissymmetry of the repulsive electrostatic potential (eqn (5)), for different values of $b$, the radius of crowding spheres. Simulations were run with the hard core model for the DNA chain and $N=500$ octahedral crowders, each octahedron being composed of six spheres of radius $b$. $R_g$ was computed after equilibration of the full system.

**Figure 6**: Plot of $R_g$, the mean radius of gyration of the DNA chain, as a function of $\delta$, the dissymmetry of the repulsive electrostatic potential (eqn (5)), for spherical (disks), dumbbell (squares), and octahedral (triangles) crowders at heavy (filled symbols) and light (empty symbols) crowding conditions. These plots are taken from Figs. 3, 4, and 5, and are superposed here for the sake of an easier comparison. The crowder volume occupancy ratio $\rho$ for each curve is indicated in the legend.

**Figure 7**: Plot, as a function of the distance $r$ between their centers, of the repulsion energy between two DNA beads (blue long-dashed curve), a DNA bead and an octahedron (green short-dashed curve), and two octahedra (red solid curve), for $e_C = e_{DNA}$, $b = 6.7$ nm, and $\delta = 0$. The geometry of octahedra is frozen at the equilibrium conformation and energy is minimized over all orientations of the octahedra, as described in section 3.2. The grey short-dashed curve, which nearly superposes on the green short-dashed one, describes the interaction energy between a DNA bead and a single crowding sphere of radius $b = 9.75$ nm. The grey solid curve, which nearly superposes on the red solid one, describes the interaction energy between two crowding spheres of radius $b = 10.75$ nm. The horizontal dot-dashed line denotes thermal energy.

**Figure 8**: (**Inset**) Plot of $p_X(r)$, the mean density distributions of DNA beads and crowding spheres, for the equilibrated system composed of the DNA chain, 1000 big crowding spheres of radius $b_B = 7.10$ nm, and 1000 smaller crowding spheres of radius $b_S = 5.77$ nm. The vertical dot-dashed line is located at $r = 60$ nm. (**Main plot**) Plot of $Q_{S/B}$, the enrichment of small crowding spheres relative to big ones in the central part of the confining sphere ($r \leq 60$ nm), as a function of $b_B / b_S$, the ratio of the radii of big and small crowding spheres. Besides

The system is composed of the DNA chain, 1000 big crowding spheres of radius $b_B$ and 1000 smaller crowding spheres of radius $b_S$.

**Figure 9** : Plot, as a function of $N_S$, the number of small crowding spheres, of $R_g$, the mean radius of gyration of the DNA chain (red disks, left axis), and $Q_{S/B}$, the enrichment of small crowding spheres relative to big ones in the central part of the confining sphere (blue lozenges, right axis), for the equilibrated system composed of the DNA chain, $N_S$ small crowding spheres of radius $b_S = 5.0$ nm, and $N_B = 400$ bigger crowding spheres of radius $b_B = 10.0$ nm.

**Figure 10** : Plot, as a function of $N_S$, the number of small crowding spheres, of $R_g$, the mean radius of gyration of the DNA chain (red disks, left axis), $Q_{S/(B+M)}$, the enrichment of small crowding spheres relative to big and medium-sized ones in the central part of the confining sphere (blue filled lozenges, right axis), and $Q_{M/B}$, the enrichment of medium-sized crowding spheres relative to big ones in the central part of the confining sphere (blue empty lozenges, right axis), for the equilibrated system composed of the DNA chain, $N_S$ small crowding spheres of radius $b_S = 5.0$ nm, $N_M = 160$ medium-sized crowding spheres of radius $b_M = 7.94$ nm, and $N_B = 320$ big crowding spheres of radius $b_B = 10.0$ nm.

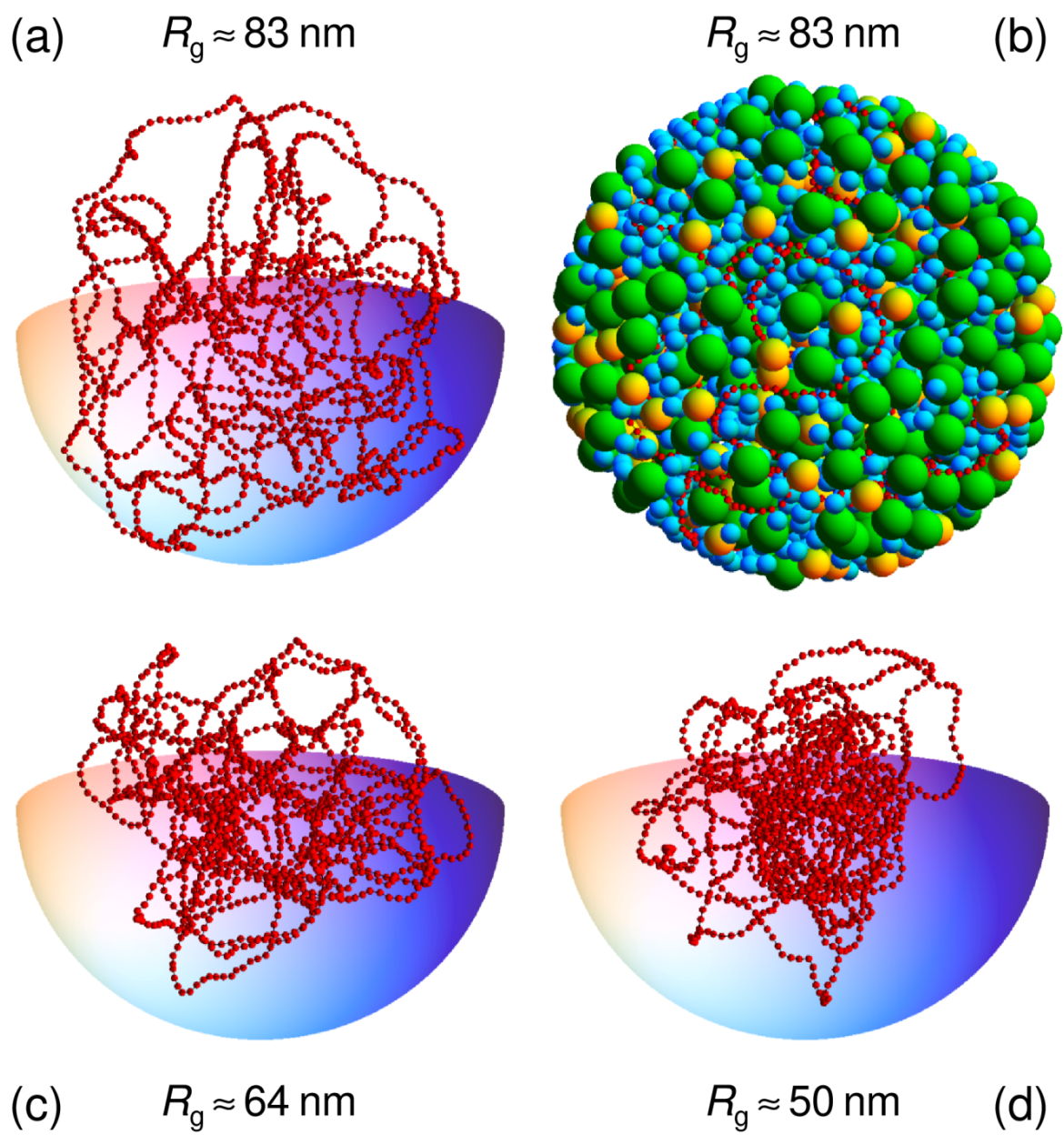

**Figure 2**

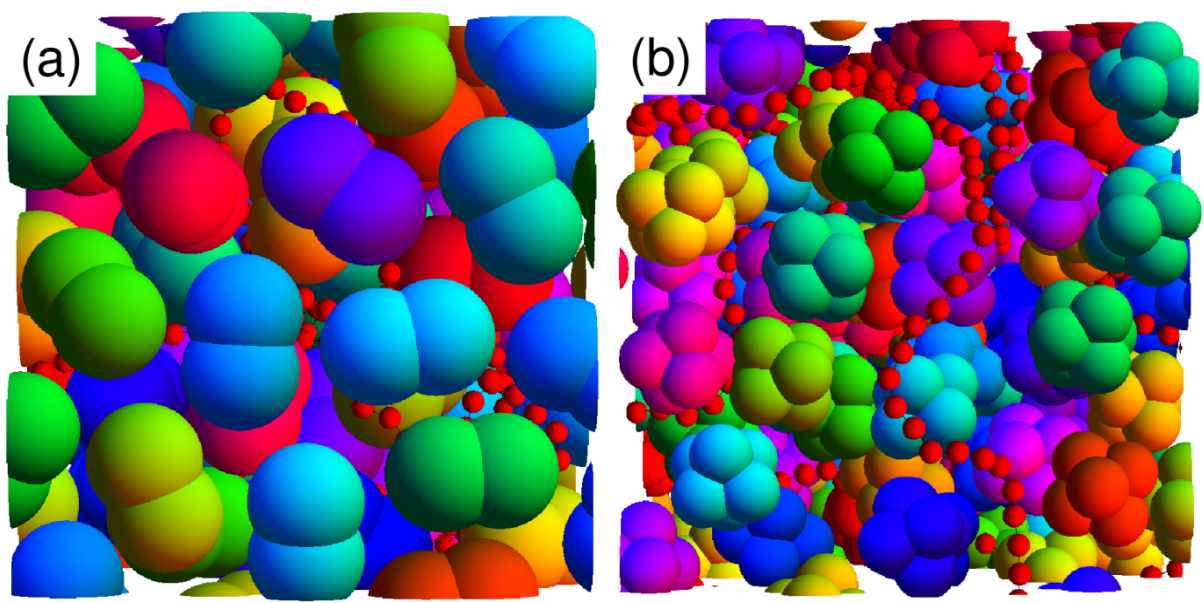

**Figure 3**

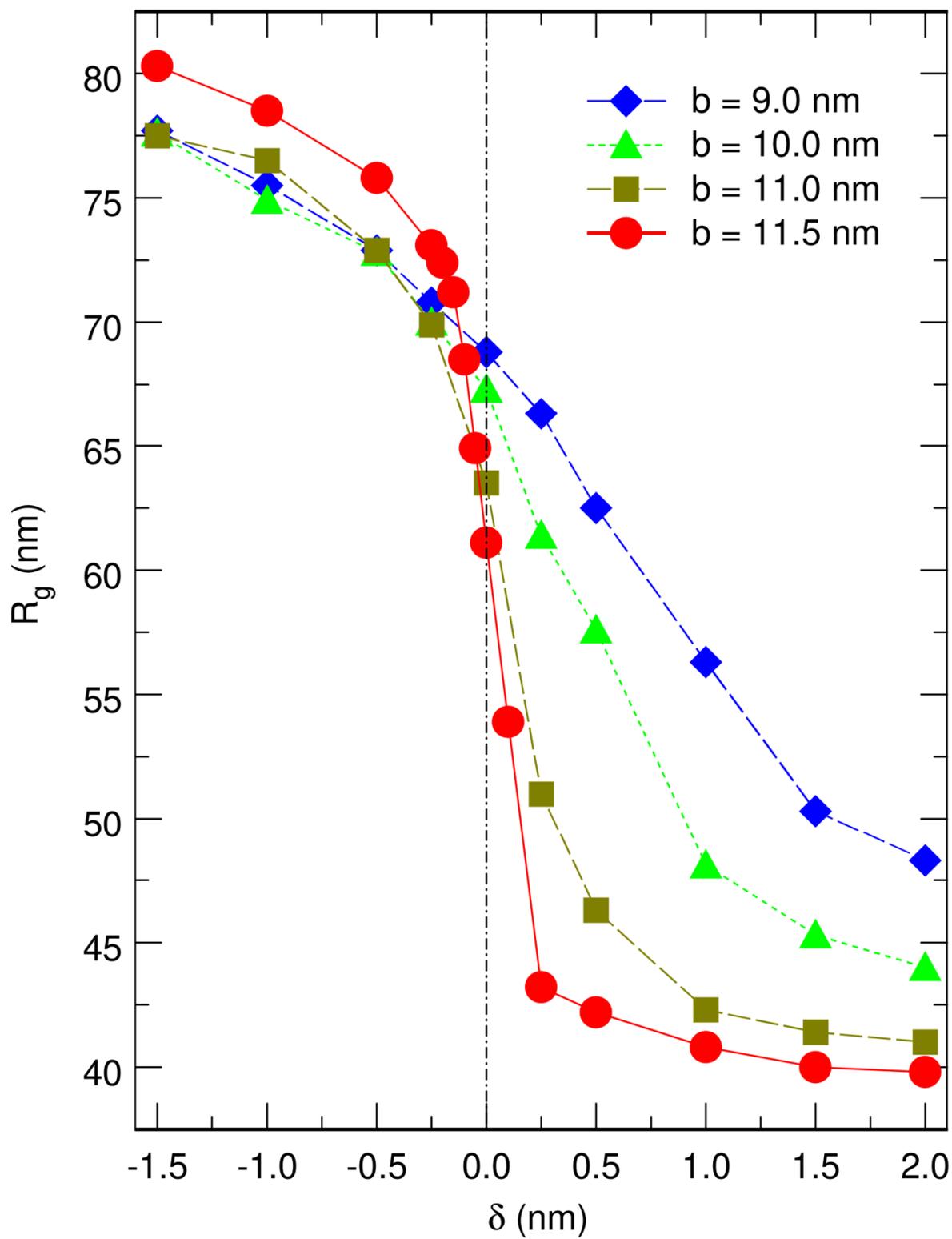

**Figure 4**

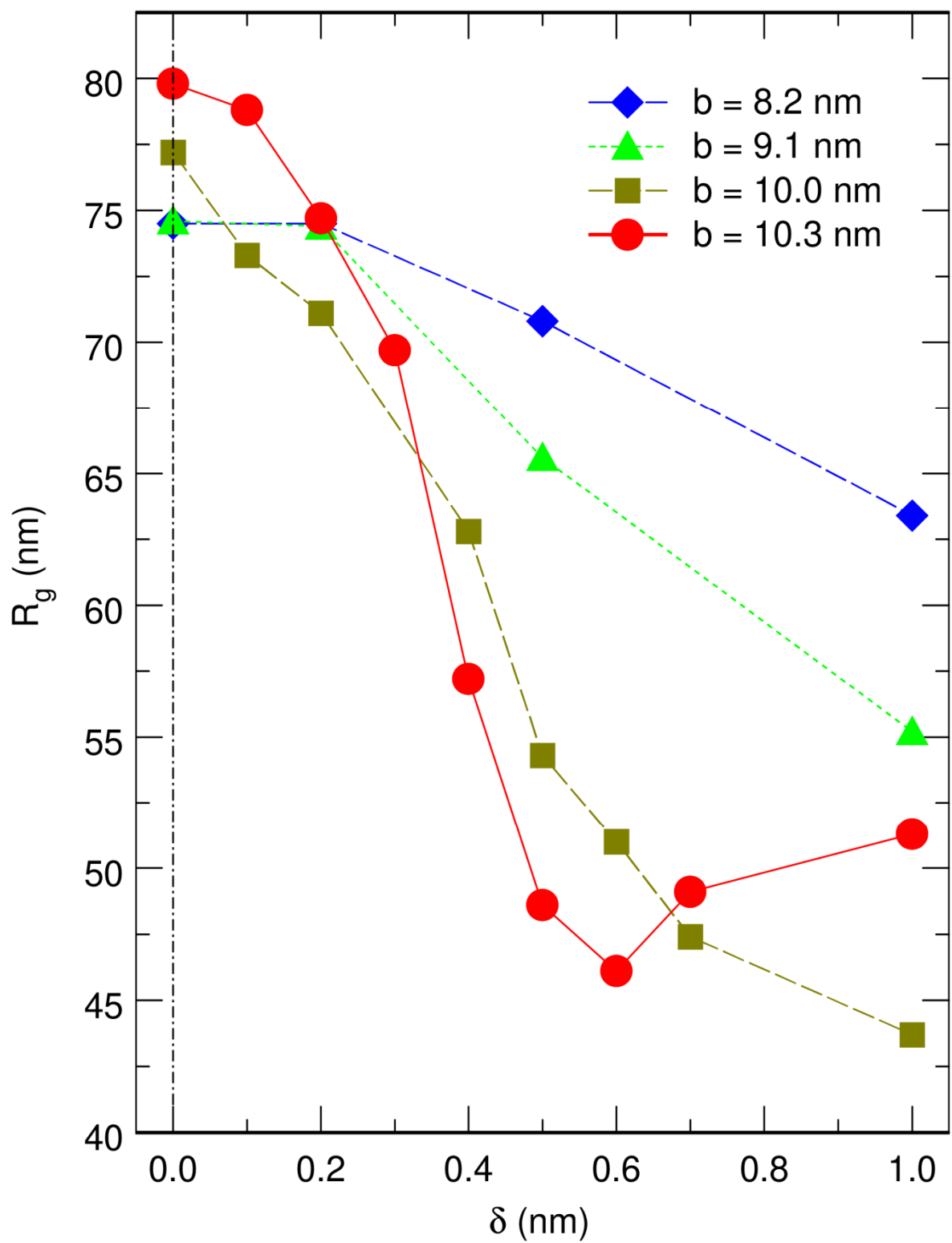

**Figure 5**

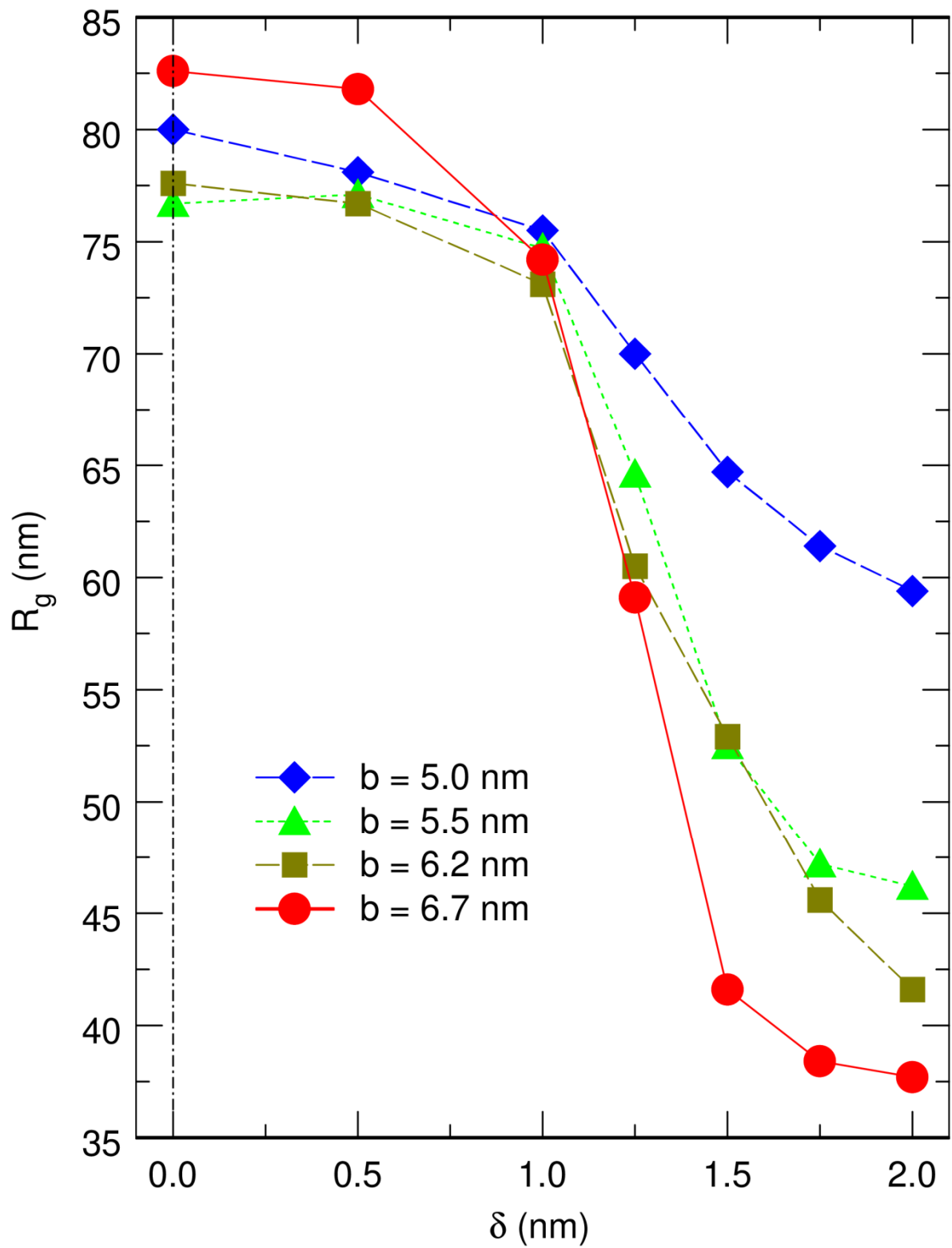

**Figure 6**

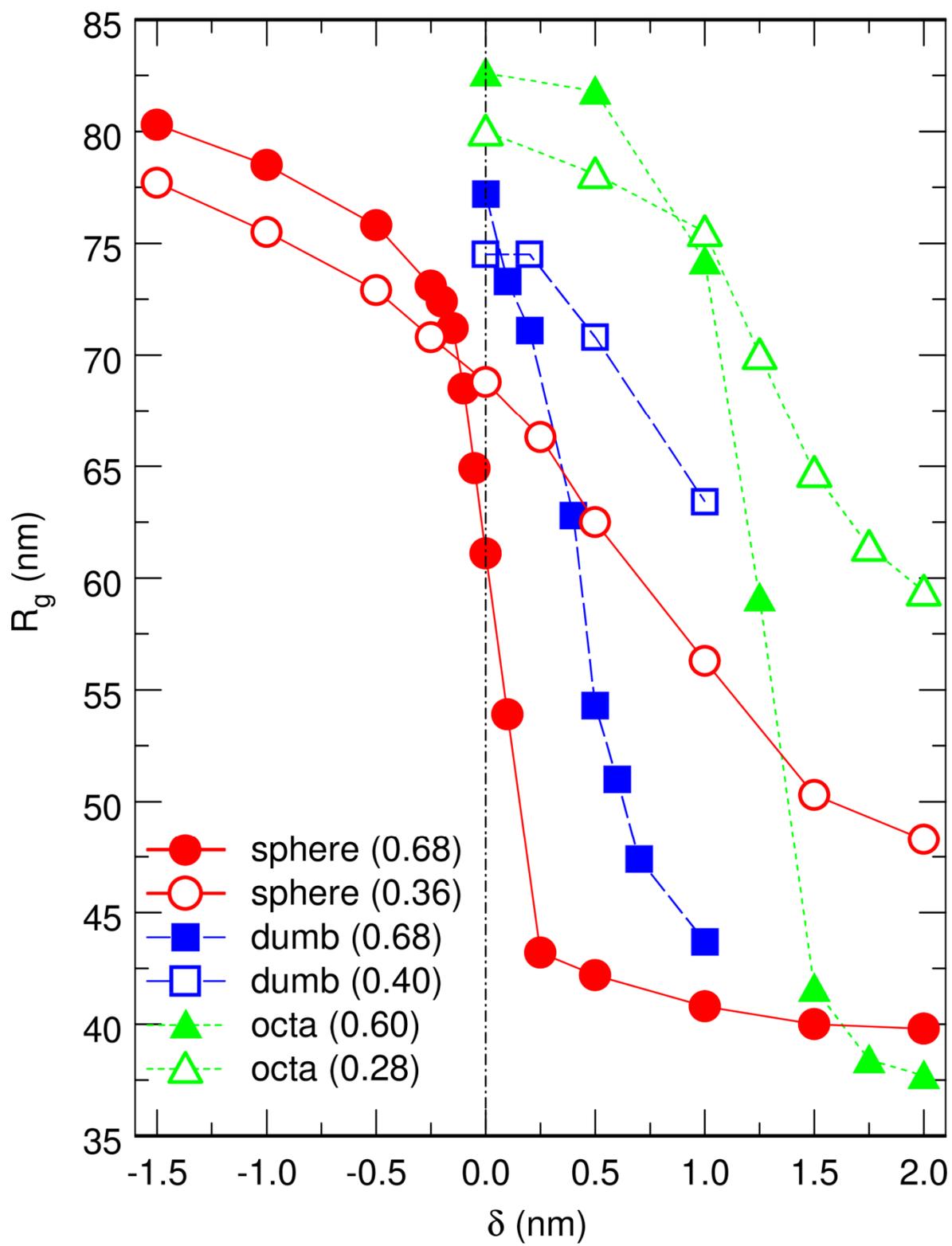

**Figure 7**

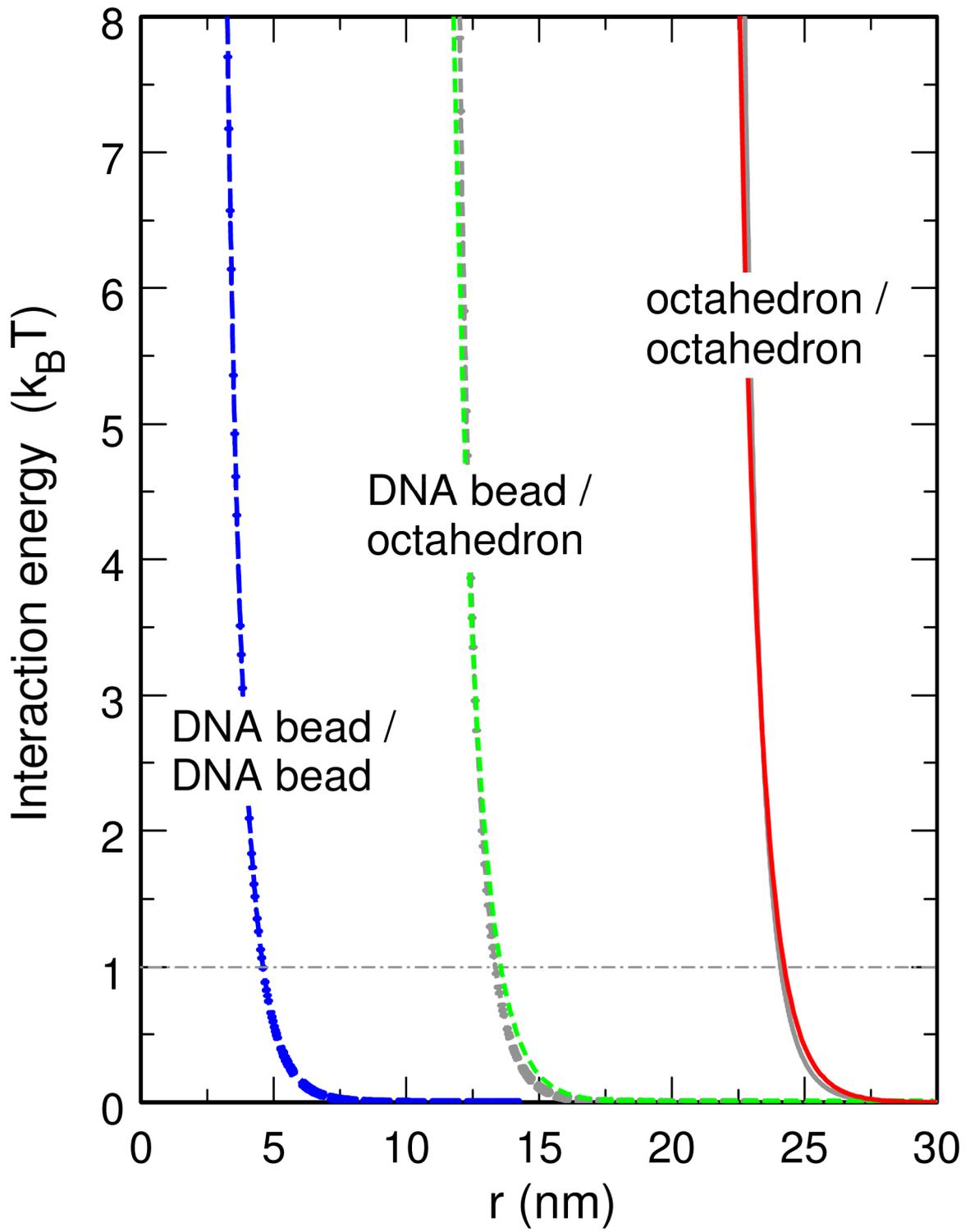

**Figure 8**

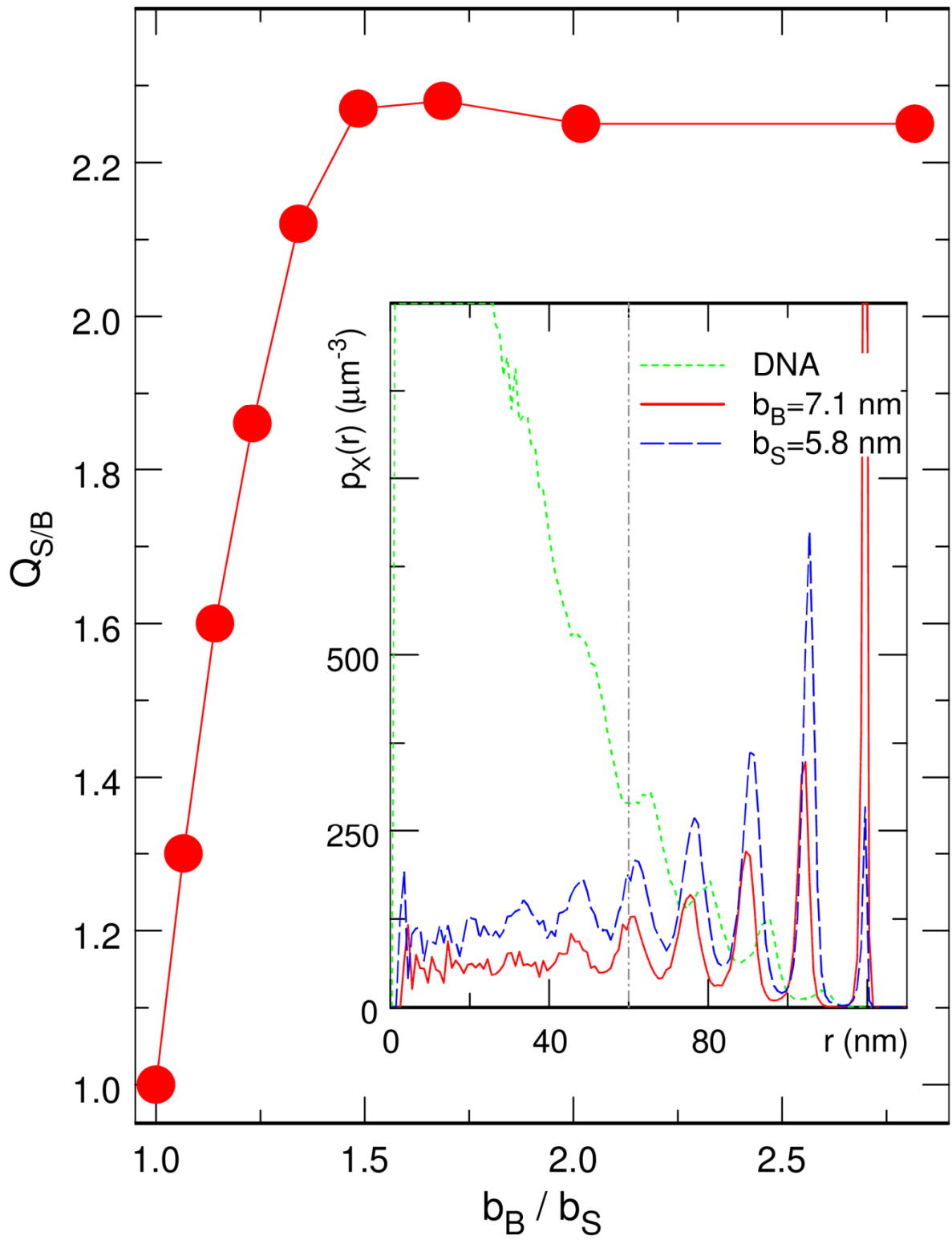

**Figure 9**

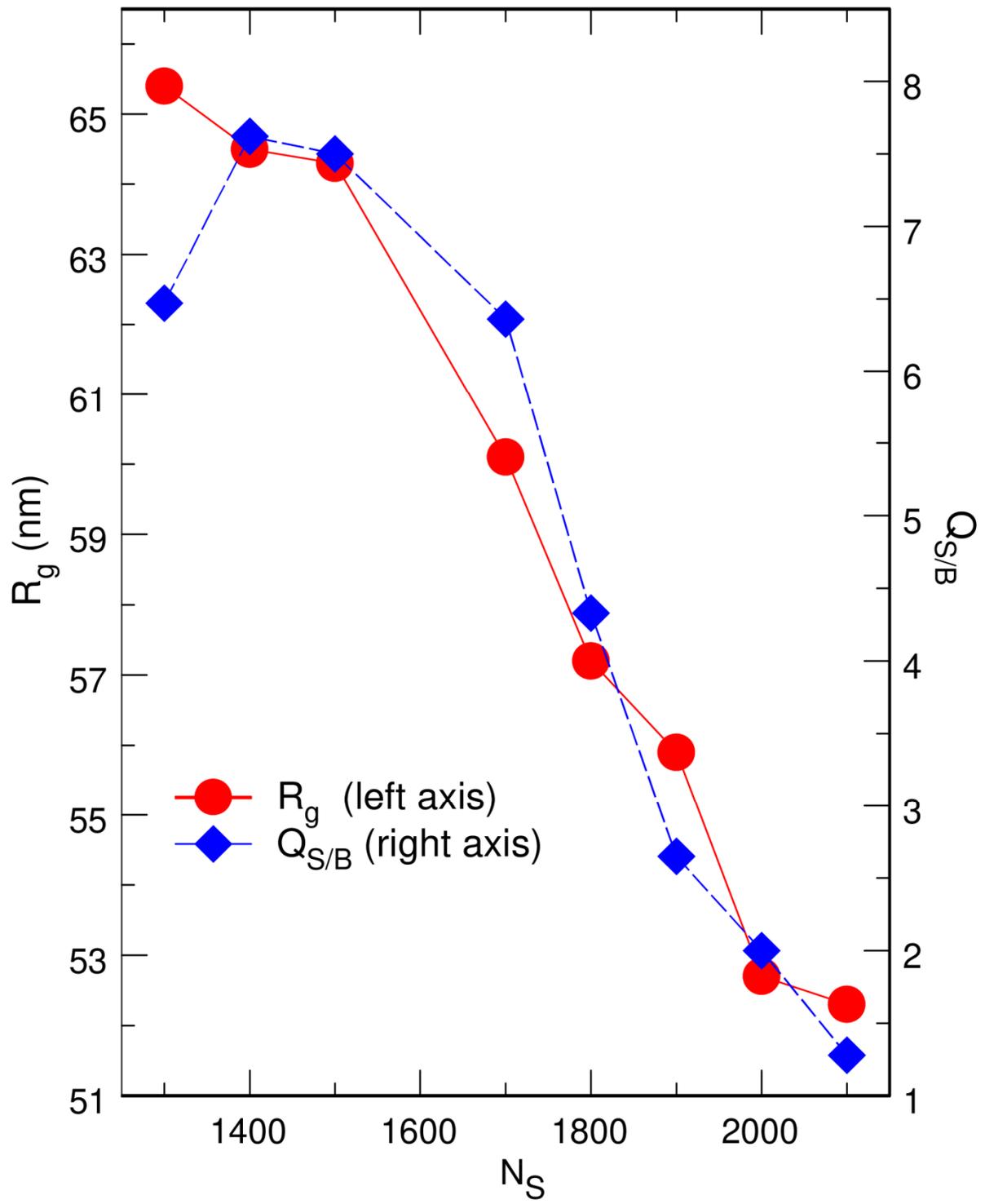

**Figure 10**

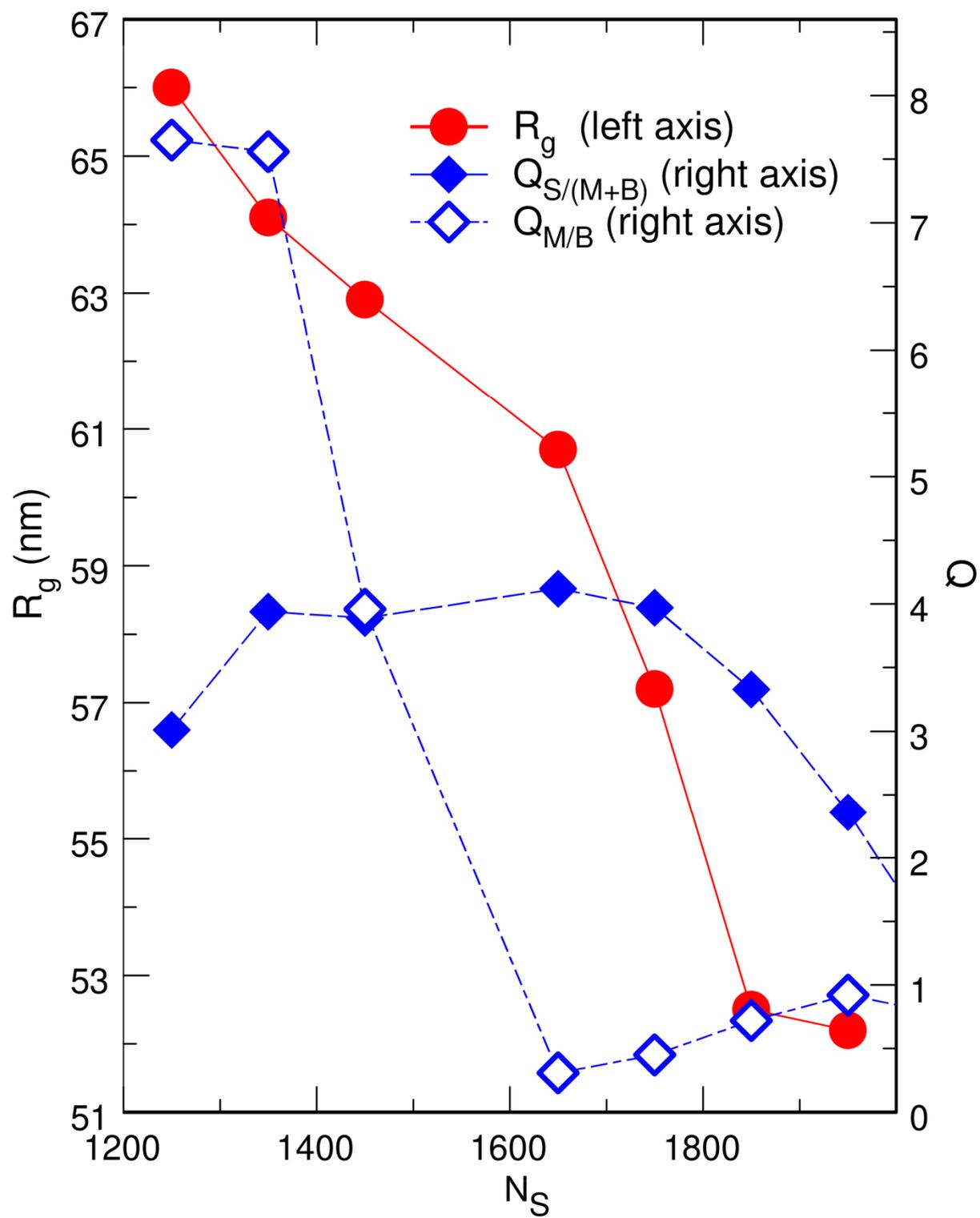